# Method of Running Sines:
# Modeling Variability in Long-Period Variables

Ivan L. Andronov[1], Lidia L. Chinarova[2]

[1] Department "High and Applied Mathematics", Odessa National Maritime University, Ukraine,
*tt_ari @ ukr.net*
[2] Astronomical Observatory, Odessa National University, Ukraine,
*lidia_chinarova @ mail.ru*

*"Even Ptolomeus knew the Fourier transform.*
*Thus no new methods for time series analysis are needed."*
*Anonymous referee, 1997*

**Abstract.**

We review one of complementary methods for time series analysis – the method of "Running Sines". "Crash tests" of the method include signals with a large period variation and with a large trend. The method is most effective for "nearly periodic" signals, which exhibit "wavy shape" with a "cycle length" varying within few dozen per cent (i.e. oscillations of low coherence). This is a typical case for brightness variations of long-period pulsating variables and resembles QPO (Quasi-Periodic Oscillations) and TPO (Transient Periodic Oscillations) in interacting binary stars – cataclysmic variables, symbiotic variables, low-mass X-Ray binaries etc. General theory of "running approximations" was described by Andronov (*1997A&AS..125..207A*), one of realizations of which is the method of "running sines". The method is related to Morlet-type wavelet analysis improved for irregularly spaced data (Andronov, *1998KFNT...14..490A*, *1999sss..conf...57A*), as well as to a classical "running mean" (="moving average"). The method is illustrated by an application to a model signal with strongly variable period, as well as to a semi-regular variable AF Cyg. Some other stars studied with this method are discussed, e.g. RU And (switching between "Mira-type" large amplitude oscillations and time intervals of "constancy"), intermediate polars MU Cam (1RXS J062518.2+733433) and BG CMi, magnetic dwarf nova DO Dra, symbiotic stars UV Aur and V1329 Cyg.

## Introduction

Many stars of different types of variability exhibit changes in all parameters of the light curve, e.g. Long-Period pulsating variables (LPVs), cataclysmic, symbiotic and X-Ray binaries. The astronomical observations are typically irregularly spaced in time due to numerous gaps. So the Fourier transform, which suggests a continuous function defined from $-\infty$ to $+\infty$, or even its "periodic simplification" (which requires continuous or regularly spaced signal defined at a time interval of length equal to the basic period), may not be applied.

For the periodogram analysis, dozens of related methods have been elaborated. They have been discussed e.g. in monographs of Pelt (1983) and Terebizh (1992). These methods may be subdivided into two large groups – "parametric" and "non-parameric" ones. The first group of methods is based on "main distance" of the observations from values of an approximating function (best fit) at the same argument (time, phase). Thus sometimes such methods are also called "point-curve" ones. Very often the main problem is to determine not only the class of basic functions (e.g. algebraic or trigonometric polynomials), but also a statistically optimal model (number of parameters). For "global" approximations (i.e. taking into account all available data $m_i, t_i$, $i=1..N$), the statistically optimal fit may be determined using the following criteria (see Andronov 1994, 2003 for detailed description):

1.      the Fischer's criterion (typically leading to extra large number of parameters as compared to other methods);

2.      the "best accuracy" criterion (with many modifications – minimal r.m.s. accuracy of the approximation function at arguments of observations or integrated over all interval; at some specific argument t; etc.).

3.      the criterion of maximal "signal/noise" ratio (especially important for variable signals, as a previous criterion sometimes may recommend a model of "constant" for a very noisy signal).

Andronov (2012ab) discussed "special shapes" for modeling periodic signals, which may significantly decrease a number of statistically optimal parameters and to improve both accuracy and "signal/noise" ratio.

The "non-parametric" methods are based on a minimization of sum of "generalized distances" between points subsequent in phase (for a given trial period). The most famous from them is the method of Lafler and Kinman (1965) improved by Kholopov (1971). Comparative analysis of 9 modifications of this method was presented by Andronov and Chinarova (1997).

These methods deal to signals assuming to be "periodic" in a sense that deviations from periodicity or changes of the shape are neglected.

**Global vs. "Running" Approximations**

The "running" (often called "local") approximations differ from the "global" ones by a variety of approximating functions which depend not only on a trial argument (typically time) $t$, but also (in a "wavelet terminology") "shift" $t_0$ and "scale" $\Delta t$ (see Andronov 1997 for details):

$$\varphi(t, t_0, \Delta t) = \sum_{\alpha=1}^{m} C_\alpha(t_0, \Delta t) \cdot f_\alpha(t - t_0) \tag{1}$$

Here $C_\alpha$ are called "coefficients" and $f_a$ – "basic functions". For a fixed set of basic functions, the coefficients may be determined e.g. using the weighted least squares and minimizing the test function

$$\Phi(t_0, \Delta t) = \sum_{k=1}^{N} p(z) \cdot w_k \cdot (x_k - \varphi(t_k, t_0, \Delta t))^2, \quad z = (t - t_0)/\Delta t \qquad (2)$$

in respect to unknown coefficients $C_\alpha$. Here there are introduced simultaneously two types of coefficients: usual ones $w_k = \sigma_0^2/\sigma_k^2$, where $\sigma_0$ is "unit weight error" and $\sigma_k$ is accuracy of $k^{th}$ observation $x_k$, and an additional weight (filter) function $p(z)$.

For "global" approximations, $p(z)=1$, i.e. no dependence of $\Phi(t_0, \Delta t)$ on these parameters. For "running" ones, $p(z)$ may be defined on all $z$, e.g. a Gaussian function $p(z)=\exp(-cz^2)$, as used in the improvement of the Morlet-type wavelet transform for irregularly spaced data (Andronov 1998, 1999).

For "local" (or "running") approximations, $p(z)>0$ only for small values of $|z|\leq 1$. The most known is the method of the "running mean", also named as "running average", "moving average", "sliding mean/average", "rolling mean/average". In this case, the signal in the interval $[t_0-\Delta t, t_0+\Delta t]$ is approximated with a constant $C_1$, i.e. the only basic function is $f_1(z)=1$ (i.e. the number of parameters $m=1$). Then the interval of smoothing is shifted, the parameters of the function $\varphi(t, t_0, \Delta t)$ are recomputed etc. However, for "unweighted" "running mean", $p(z)=1$, if $|z|\leq 1$.

For "weighted" "running mean" ("moving average" etc.), the weights $p(z)$ are generally dependent on $z$, but still $m=1$. A number of filters are discussed by Hamming (1977).

The function defined in Eq. (1) depends on 3 variables (in fact, also on basic and weight functions, assuming fixed set of observations). For practical purposes, one should prefer to have a "best" smoothing function $\theta(t)$, which depends only on one parameter $t$. So there is a freedom in choosing "what is the best", and so there are numerous approaches.

The classical approach is to decrease the number of parameters by fixing some of them, typically

$$\theta(t_0) = \varphi(t_0, t_0, \Delta t) \qquad (3)$$

i.e. from an infinite number of values of the function $\varphi(t, t_0, \Delta t)$, we choose only one value, where trial time $t$ coincides with the "shift" $t_0$ (typically, center of the interval of smoothing). The parameter $\Delta t$ is fixed in some way ("user choice" based e.g. on spectral properties of the signal, as smoothing acts like a low-frequency filter, see textbooks e.g. by Whittaker and Robinson 1944, Hamming (1977)).

Absolute majority of methods assume regularly spaced signals, defined at a grid $t_k=t_1+(k-1)\delta t$. The detailed formulae for a general case of *irregularly* spaced data were presented by Andronov (1997). Andronov (1998, 1999) also introduced a "wavelet smoothing", where, for a given trial $t_0$, the parameter $\Delta t$ does not assumed to be constant, but is determined in a statistically optimal way.

Depending on the aim of approximation, the weight functions may be symmetric ($p(-z)=p(z)$) or asymmetric. Symmetric weight functions are typically used when the data have been already obtained previously, and we smooth the values. Asymmetric ones are typically used for prediction purposes, when only "past" values are known (i.e. $t_k < t_0$ for all $k = 1..N$).

Thus formally the coefficients and basic functions in Eq. (1) may be interchanged:

$$\theta(t_0, \Delta t) = \sum_{\alpha=1}^{m} C_\alpha(t_0, \Delta t) \cdot f_\alpha(0) \tag{4}$$

It should be noted that there is a class of functions which may produce the same functions $\theta(t_0) = \varphi(t, t_0, \Delta t)$ for all $t_0$ (for continuous or exactly defined signals without errors of measurements). The necessary condition for such functions is the possibility of computations for all $z$ and $\lambda$:

$$f_\alpha(z + \lambda) = \sum_{\beta=1}^{m} A_{\alpha\beta}(\lambda) \cdot f_\beta(z) \tag{5}$$

Here $A_{\alpha\beta}(\lambda)$ is a $m \times m$ matrix. A common example is $f_1(z) = 1$, $f_\beta(z) = z^k \cdot \exp(\gamma z) \cdot \cos(\eta z)$, $f_{\beta+1}(z) = z^k \cdot \exp(\gamma z) \cdot \sin(\eta z)$, $k = 1..p$, where $p$ is a degree of a polynomial, and pairs of basic functions containing sine and cosine for each $\eta > 0$.

For discrete signals with rounding errors and especially measurement errors, the functions $\theta(t_0)$ and $\varphi(t, t_0, \Delta t)$ should coincide within statistical errors.

If Eq. (5) can't be satisfied, only one point $\varphi(t_0, t_0, \Delta t)$ will coincide with $\theta(t_0)$. This means that partial derivatives of the function $\varphi(t, t_0, \Delta t)$ with respect to variables $t, t_0$ and $d\theta(t_0)/dt_0$ are generally different (see Andronov 1997 for details).

In practice, often are used *separately* polynomials, exponents and sines.

## The method of "Running Sines"

For this method, $f_1(z) = 1$, $f_1(z) = \cos(\eta z)$, $f_2(z) = \sin(\eta z)$, $p(z) = 1$ (for $-1 \leq z \leq 1$), $\eta = 2\pi\Delta t/P$, $P$ is trial period. Going back from the dimensionless variable $z$, which is common in the wavelet analysis, to the argument $t$ (time in time series analysis, wavelength in the spectrum etc), the Eq.(1) may be rewritten as

$$\varphi(t, t_0, \Delta t) = C_1 + C_2 \cdot \cos(\omega(t-T_0)) + C_3 \cdot \sin(\omega(t-T_0)) =$$
$$= a - r \cdot \cos(\omega(t-T_0) - 2\pi\phi) = \tag{6}$$
$$= a - r \cdot \cos(\omega(t-T_M)) = a + r \cdot \cos(\omega(t-T_m))$$

where $\omega = 2\pi/P$, $T_0$ is called "initial epoch", $T_M$ – "epoch of maximum" (="time of maximum"="maximum timing"), $T_m$ – "epoch of minimum", $r$ – "semi-amplitude", $\phi$ – "phase" (="phase of maximum"). The parameter $a$ has a meaning of a "mean value of the smoothing function over a complete period" and is generally different from a sample mean of the signal values in the subinterval $[t_0 - \Delta t, t_0 + \Delta t]$. Other note is that the signal in astronomy is often "brightness in stellar magnitudes", i.e. the "maximum" ("Max", "M") corresponding to "maximal brightness" corresponds to "minimal stellar magnitude". This is why

Obvious relations are

$$C_2 = -r \cdot \cos(2\pi\phi)$$
$$C_3 = -r \cdot \sin(2\pi\phi)$$
$$r = (C_2{}^2 + C_3{}^2)^{1/2} \tag{7}$$
$$2\pi\phi = \arctan2(C_3, C_2) = \arctan(C_3/C_2) - \pi \cdot ((1\text{-sign}(C_2))/2) \cdot \text{sign}(C_3)$$

The last function (=atan2($C_3$;$C_2$) in the designations of electronic tables – MS Excel, GNUmeric, Open/Libre Office Calc etc.) produces a result in a range $[-\pi/2, \pi/2)$, i.e. -0.5$\leq\phi$<0.5. The maximum time $T_M$=$T_0$+$PE$+$P\cdot\phi$ may be computed for each cycle number $E$. However, for an approximation of the data in the interval $[t_0-\Delta t, t_0+\Delta t]$, it is natural to determine $T_M$ in the same interval (and corresponding $E$). For this purpose, the additional parameter $\phi_M$ is to be determined, i.e. the phase $\zeta$ corresponding to the moment $t_0$: $E_0$=$\zeta-$int($\zeta$+0.5), $\zeta$=($t_0$-$T_0$)/$P$, $\phi_0$=$\zeta-E_0$, $\psi$=$\phi-\phi_0$. Finally, if $|\psi|\leq$0.5, then $E$=$E_0$, $T_M$=$T_0$+$PE$+$P\phi$ =$t_0$+ $P\psi$. If $\psi$< $-0.5$ then one should add 1 to $E$ and $P$ to $T_M$ computed using these formulae, and subtract 1, if $\psi$> +0.5.

We remind that all parameters in Eq.(7) are functions of $t_0$ and $\Delta t$. Obviously, one should compute a covariation matrix of the deviations of the coefficients:

$$\langle C_{d\alpha} C_{d\beta} \rangle = \sigma_0^2 R_{\alpha\beta} \tag{8}$$

In a simplest case of a rectangular filter $p(z)$=1, $R_{\alpha\beta} = A_{\alpha\beta}^{-1}$, (Whittaker and Robinson 1944). For arbitrary $p(z)$, the expressions are much more complicated and analyzed in detail by Andronov (1997). They are used for "running parabola" and "wavelet" approximations.

For "running sines", we adopt $p(z)$=1, and final expressions are: $f_1$(t)=1, $f_2$(t)=cos($\omega\cdot(t$-$T_0$)), $f_3$(t)=sin($\omega\cdot(t$-$T_0$)), $m$=3,

$$A_{\alpha\beta} = \sum_{k=N_1}^{N_2} w_k f_\alpha(t_k) f_\beta(t_k) , \tag{9}$$

$$B_\beta = \sum_{k=N_1}^{N_2} w_k x_k f_\beta(t_k) , \tag{10}$$

$$C_\alpha = \sum_{\beta=1}^{m} A_{\alpha\beta}^{-1} B_\beta , \tag{11}$$

$$\sigma_0^2 = \frac{1}{N_2 - N_1 + 1 - m} \sum_{k=N_1}^{N_2} w_k \cdot (x_k - \varphi(t_k, t_0, \Delta t))^2 , \tag{12}$$

$$\sigma^2[\varphi(t, t_0, \Delta t)] = \sigma_0^2 \sum_{\beta=1}^{m} R_{\alpha\beta} f_\alpha(t) f_\beta(t) , \tag{13}$$

$$\sigma^2[\theta(t_0)] = \sigma_0^2 \sum_{\beta=1}^{m} R_{\alpha\beta} f_\alpha(t_0) f_\beta(t_0) , \tag{14}$$

$$\sigma^2[C_\alpha] = \sigma_0^2 R_{\alpha\alpha}, \tag{15}$$

$$\sigma^2[r] = \frac{\sigma_0^2}{r^2}\left[R_{22}C_2^2 + 2R_{23}C_2C_3 + R_{33}C_3^2\right], \tag{16}$$

$$\sigma^2[\phi] = \frac{\sigma_0^2}{4\pi^2 r^4}\left[R_{22}C_3^2 - 2R_{23}C_2C_3 + R_{33}C_2^2\right], \tag{17}$$

$$\sigma[T_M] = \sigma[T_m] = P \cdot \sigma[\phi]. \tag{18}$$

Here $\sigma$["parameter"] is the statistical error (accuracy) of the "parameter", $A_{\alpha\beta}$ is the matrix of normal equations, $A_{\alpha\beta}^{-1}$ is an inverse matrix, $N_1$ and $N_2$ – numbers of the first and last observational points in the time interval from $(t_0-\Delta t)$ to $(t_0+\Delta t)$.

Obviously, the sinusoidal signal $s(t)$=a-r·cos$(2\pi(t-T_M)/P)$ will be fitted exactly, i.e. $\varphi(t,t_0,\Delta t)=\theta(t_0)=s(t)$ for all $t,t_0,\Delta t$, if there are at least 3 points in the interval $[t_0-\Delta t, t_0+\Delta t]$. For observations with errors, the parameters of the fit should show scatter of a r.m.s. value corresponding to an accuracy $\sigma$["parameter"] as described above.

**"Crash Test" of the Method. I. Sample Signal with Variable Period**

A review on different definitions (continuous and discrete) of the period for study of its possible variations was presented by Kopal and Kurtz (1957). They are based on extensions of the linear ephemeris

$$T(E)=T_0+P_0 \cdot E \tag{19}$$

to a case of generally variable period, where $T$ is time, $E$ – cycle number, $T_0$ – initial epoch and $P_0$ – initial period. Following Argelander (1859), one may obviously write $P(E)$=d$T$/d$E$ for continuous case, while classically the "O – C" analysis deals with integer $E$:

$$\text{O} - \text{C} = T_E - T(E), \tag{20}$$

where $T_E$ (="O") is "Observed" moment of characteristic event (e.g. minimum of eclipsing binary, maximum of pulsating variable, crossing of $\gamma$– velocity by radial velocity etc.) and $T(E)$ (="C") is "Calculated" using Eq. (19) or (generally) other theoretical model. One may also write d$E$=d$T$/$P$, or, taking into account intrinsic frequency $f(T)$=1/$P(T)$,

$$E(T) = E_1 + \int_{T_1}^{T} f(t)\mathrm{d}t = E_1 + \int_{T_1}^{T} \frac{\mathrm{d}t}{P(t)} \tag{21}$$

Similar definitions of this kind ("instantaneous phase") are widely used in signal processing (cf. Cohen 1995). Application of this formalism for variable stars is presented by Mikulášek et al (2012).

Typically period variations are slow and $T(E)$ is modeled (e.g. Tsessevich 1970). An atlas of "O–C" diagrams for hundreds of eclipsing binaries was published by Kreiner, Kim and Nha (2001), which show long-term ("secular") period increase and decrease, periodic variations due to light-time effect (discovered by Ole Rømer in 1676) or apsidal motion (cf. Sterne 1939).

For our "crash test", we study a model signal defined at 1001 points $t_k = 0..1000$ with a period changing with time at a constant rate $\dot{P} = (P_2 - P_1)/(T_2 - T_1)$:

$$P(T) = P_1 + \dot{P} \cdot (T - T_1) \qquad (22)$$

From Eq. (21) and (22) one gets

$$E(T) = E_1 + \frac{1}{\dot{P}} \cdot \ln\left(1 + \dot{P} \cdot \frac{T - T_1}{P_1}\right) = E_1 + \frac{1}{\dot{P}} \cdot \ln\left(1 + \dot{P} \cdot E_0\right) \approx$$

$$\approx E_1 + E_0 - \dot{P} E_0^2 / 2 + \dot{P}^2 E_0^3 / 3 + ... + (-1)^{k-1} \dot{P}^{k-1} E_0^k / k + ... \qquad (23)$$

and an inverse relation

$$T(E) = T_1 + \frac{P_1}{\dot{P}} \cdot \left(e^{\dot{P} \cdot (E - E_1)} - 1\right) = T_1 + \frac{P_1}{\dot{P}} \cdot \left(e^{\dot{P} \cdot E_3} - 1\right) \approx$$

$$\approx T_1 + P_1 E_3 + P_1 \dot{P} E_3^2 / 2! + P_1 \dot{P}^2 E_3^3 / 3! + ... + P_1 \dot{P}^{k-1} E_3^k / k! + ... \qquad (24)$$

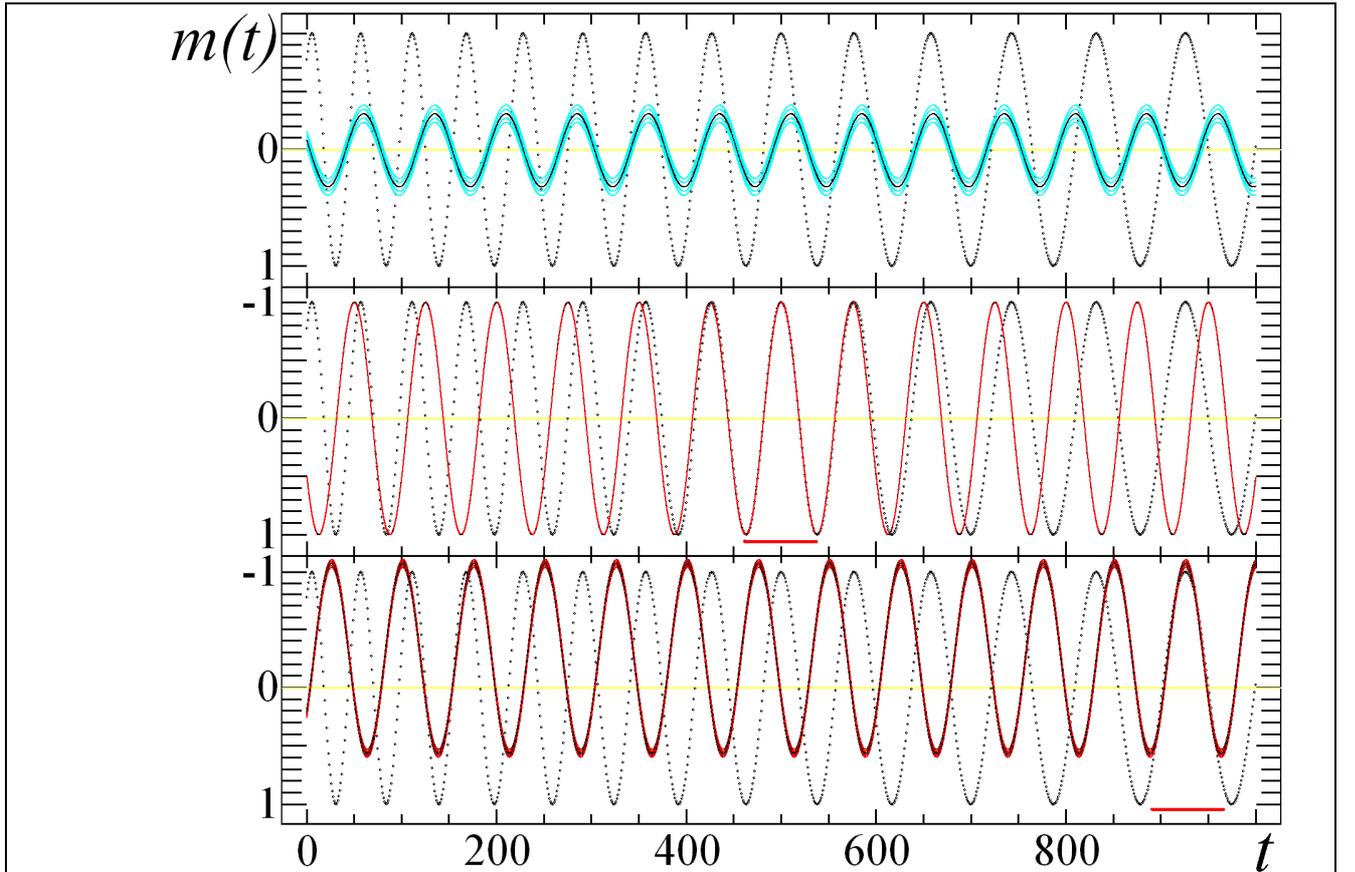

Fig. 1. The signal with variable period $s(t)$ – points, its "sine" approximations (Eq. (3)) – "global" (up) and "local" (middle, bottom), for which the interval of smoothing is marked with red segment. Red line shows fits $\varphi(t, t_0, \Delta t)$ with $\pm 1\sigma$ and $\pm 2\sigma$ error corridors.

Here $E_0=(T-T_1)/P_1$, $E_3=E-E_1$. These values coincide in a case of constant period, but are different in a general case.

For illustration, we used $P_1=75$, $T_1=500$, $E_1=0$, $P_2=100$, $T_2=1000$, i.e. the period varies from $P_1=50$ at $T=0$ to $P_2=100$ at $T=1000$. The signal is determined as

$$s(T)=0-1\cdot\cos(2\pi(E(T)-E_1)),\qquad(25)$$

where $E(T)$ is defined by Eq. (23). Here "0" and "1" are written to mark theoretical parameters $(a, r)$ of the model signal.

For comparison of the methods, we show a "global" approximation of the signal (assuming a constant period $P_1$) as in Eq.(6). One should mention a bad coincidence of the approximation with the initial signal. This is expected for such a significant period variation. In Fig.2. we show a periodogram $S(f)$ (Andronov 1994, 2003). One may note that the peak at the periodogram is wide, which corresponds to a range of variability of the period from 50 to 100. The *mean* value of the *period* $(P_1+P_2)/2=75$ corresponds to a local minimum, so the approximation is bad. The period corresponding to a *mean frequency* $(E_2-E_1)/(T_2-T_1)$ is only slightly smaller (72.13). However, the highest peak is at 84.32 and does not coincide with any of the above periods. Numerical experiments (with other duration of the interval of data, however, with the same $P_1$ and $P_2$) show a highest peak at longer periods, i.e. becoming (with an increasing length $T_2-T_1$) asymptotically closer to a larger limiting period. Also there are many apparent peaks towards shorter period. This may be explained by slower phase changes at longer periods.

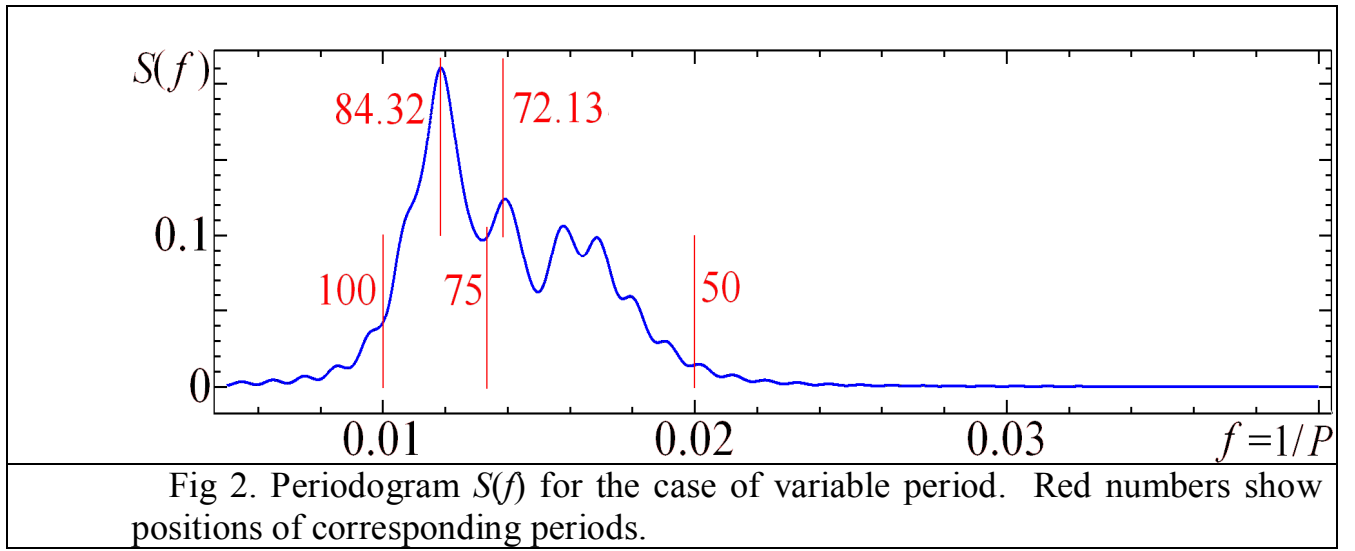

Fig 2. Periodogram $S(f)$ for the case of variable period. Red numbers show positions of corresponding periods.

The formal error estimates of the smoothing function (as in Eq. (13)) are small because the number of generated points ($N=1001$) is large (they decrease proportionally to $\sim N^{-1/2}$). This apparent contradiction is due to a suggestion that the deviations of the signal values from the fit are not correlated. An improved version of the least squares taking into account that the covariation matrix of the errors of observations is not diagonal is discussed by Andronov (1997).

The "local" fits are also shown for example for two values of $t_0$. The "filter half-width" was chosen to be $\Delta t = P_0/2$, i.e. the full width is $2\Delta t = P_0$. For $t_0 = 500$ (when the intrinsic period is close to $P_0$), the coincidence is good in the interval of smoothing $[t_0 - \Delta t, t_0 + \Delta t]$ and the difference increases with either increasing, or decreasing $t_0$. For the "right" interval, the intrinsic period $P$ is much larger than $P_0$. In this case, the "best fit" amplitude is larger than the "true" unity, the coincidence in the middle of the interval of smoothing is better than that at the borders.

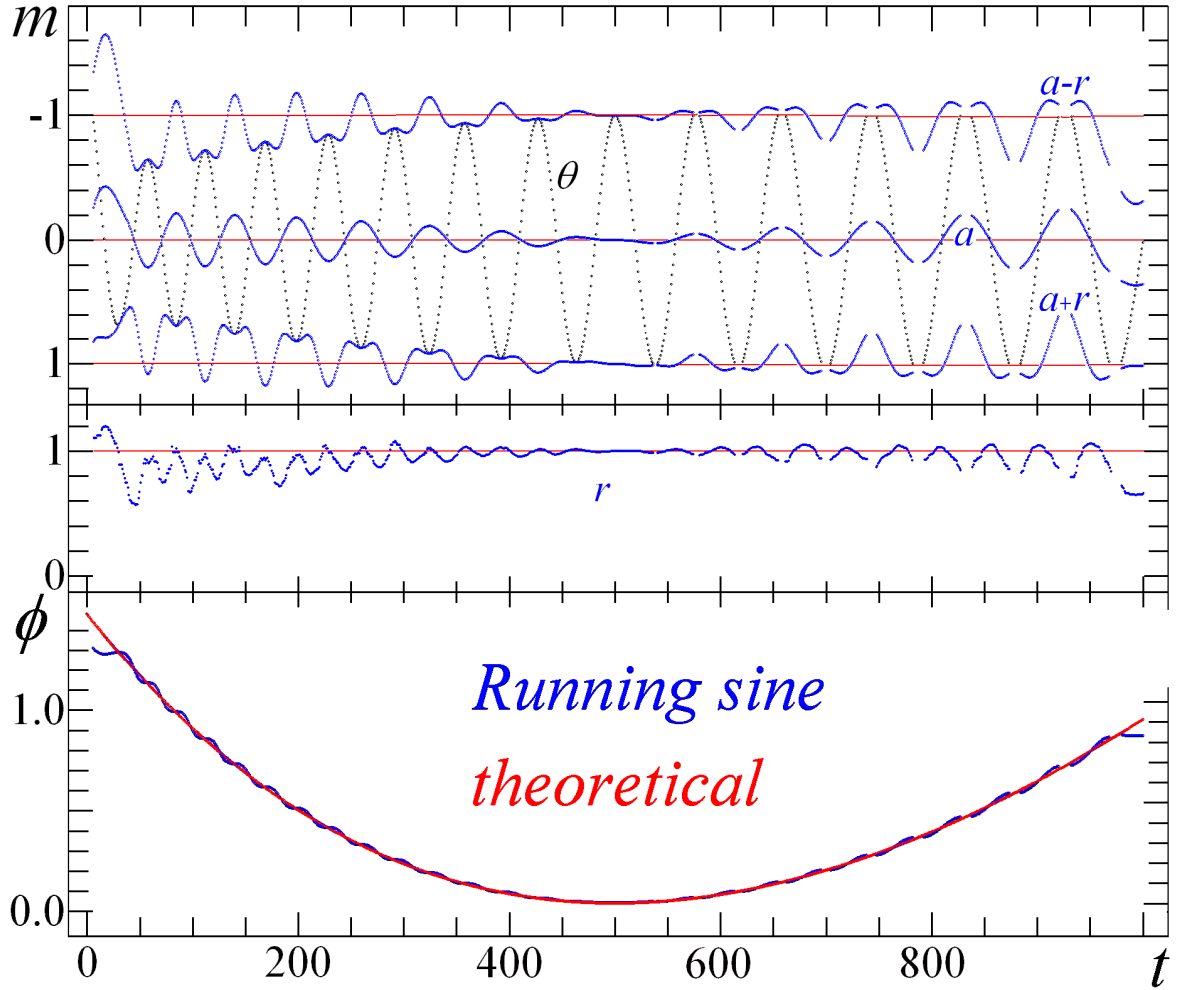

Fig. 3. Characteristics of the "Running Sine" approximation (blue) and theoretically defined values (red).

In Fig. 3 we compare results of the "Running sine" approximations with that expected from the properties of our model signal. The period is variable. However, the amplitude is constant, as well as the values of maximum ($a-r$) and minimum ($a+r$) brightness. In the approximation, the function $\theta(t_0, \Delta t)$ systematically decreases in amplitude towards shorter periods and reaches 80% of the original amplitude at $P \sim (0.75 \div 0.80) P_0$ (slightly different for different length of signal, 0.786 for a "continuous" case). For increasing $P > P_0$, the amplitude increases by a factor, which is equal to 1 at $P=P_0$, monotonically increases to 1.075 at $P=1.48P_0$ and then asymptotically decreases to 1.

The parameter $a=C_1$ formally varies with time, having maxima "in phase" with $\theta(t_0,\Delta t)$ for $P>P_0$, and in "opposite phase" for $P>P_0$. The semi-amplitude $r$ shows a "double frequency" (or "two waves"), and the maxima are close in time either to maxima, or minima of $\theta(t_0,\Delta t)$ for $P<P_0$. For $P>P_0$, the maxima of $r$ occur close to a middle of ascending and descending branches.

The phase changes slightly resemble a parabola. Practically, the parabolic fits are most commonly used for studying "period variations at a constant rate", cf. Tsesevich (1970), Kreiner, Kim and Nha (2001)). However, correct theoretical values of $E(T)$ for each time may be determined using Eq.(23), and

$$\zeta=E_0+E_1-E(T), \qquad (26)$$
$$\phi=\zeta-\text{int}(\zeta+0.5). \qquad (27)$$

Here we suggest that "initial" phase $E_1$ at $T=T_1$ may be generally different from zero. However, the formula may be slightly simplified by setting $E_1=0$, e.g. using $T_0=T_1$, $P_0=P_1$.

As usual, the resulting $\phi$ is in the interval $[-0.5,+0.5)$. However, to avoid jumps by an integer number, one should use an extended version of definition of phase, taking into account that one may add an arbitrary integer number to $\phi$, and the result will have the same sense. This is why the phase at Fig.3 varies from 1.8 instead of "from -0.2 to -0.5, then a jump to +0.5 and continuation of decrease to -0.5 etc." Such "jumps" may be removed by adding/subtracting 1 e.g. in the software MCV (Andronov and Baklanov 2004).

The phase computed using the "running sine" approximation is in an excellent agreement with theory, nearly within thickness of line, except intervals close to borders (where the observations fill only a part of the interval of smoothing). Moreover, there is an excellent coincidence of phases at the maxima and minima of $\theta(t_0,\Delta t)$.

Under strict limitations, one may suggest to determine the parameters of the approximation only in the vicinities of maxima and minima. In this case, all the parameters ($a$, $r$, $a$-$r$, $a$+$r$, $\phi$) will be changing with time smoothly, as all "oscillations" will not be seen at a graph. This is preferable "in theory", but real observations often have gaps, also close to maxima or minima, thus we have to study behavior of model parameters and functions without these limitations.

Of course, the best fit will be for $P{\approx}P_0$, but we specially made a "crash test" of very large period variations to see the range of periods $P$ (for a given $P_0$), where the method is effective.

**"Crash Test" of the Method. II. Sample Signal with Significant Trend**

Another challenge for the models is a presence of trends in an addition to oscillations with more or less stable period. Typically such type of variability is present in dwarf novae of SU UMa-type (UGSU), which show superhumps during superoutbursts, some semi-regular pulsating stars etc. For the "crash test", we assume a model with a linear trend and coherent oscillations with a period $P$:

$$s(T)=A+B \cdot (T-T_0) - R \cdot \cos(\omega(T-T_0)), \tag{28}$$

where $\omega=2\pi/P$. As the oscillations repeat, no new information is expected for many cycles, as it was for a case of variable period. Thus we have used again 1001 points $t_k=0..1000$, $P=250$, $A=5$, $B=0.008$, $R=1$. Contrary to Eq. (6), one may mention that the parameter $a(T)=A+B \cdot (T-T_0)$ in Eq.(28) is variable, and $B=da/dT$. We may introduce a related *dimensionless* parameter

$$D= da/dT \cdot P/R = d(a/R)/(dT/P) , \tag{29}$$

which shows, how much changed a parameter $a$ (in units of semi-amplitude $R$) during one period. To obtain theoretical moments of maximum, we have to solve an equation $ds(T)/dT=0$, which is, for our model, equal to

$$ds(T)/dT= B + R \cdot \omega \cdot \sin(\omega(T-T_0)) =$$
$$R \cdot \omega \cdot (D/2\pi +\sin(\omega(T-T_0)) = R \cdot \omega \cdot (D/2\pi +\sin(2\pi E_0)) \tag{30}$$

For the extrema, $E_M= -(\arcsin(D/2\pi))/2\pi+k$ and $E_m=+(\arcsin(D/2\pi))/2\pi+k+1/2$ for maxima (i.e. minima of stellar magnitude) and minima, respectively, where $k$ is an integer. Thus two extrema per period may be seen, if $|D|<2\pi$, or $|B|<2\pi R/P$. Otherwise the trend prevents extrema to occur.

In Fig. 4, is shown an illustrative function with a large value $D=2$ with a corresponding phase shift $\phi= -(\arcsin(D/2\pi))/2\pi = -0.0516$. The "global" sine fit (6) is definitely a bad approximation showing (obviously) no trend, but also a large phase shift from either function $s(T)$ (Eq.(28)), or its "de-trended" part

$$s_0(T)=A - R \cdot \cos(\omega(T-T_0)). \tag{31}$$

It should be noted that, even if $R=0$, but $B\neq0$, the "running sine" fit will lead to non-zero value of semi-amplitude $R_B=BP/\pi$ (in a case of "continuous" signal within interval of smoothing).

The shift of the position of extremum of the signal in a presence of trend is obvious from the mathematical point of view, as the root $t_e$ of the equation

$$d(s_0(t)+ B \cdot (t-T_0))/dt=0 \tag{32}$$

is distinctly different from the root $t_{e0}$ of the equation $ds_0(t)/dt=0$. For twice differentiable function $s_0(t)$,

$$\delta =t_e-t_{e0} = B/(d^2s_0(t_{e0}+\lambda\delta)/dt^2) \approx B/(d^2s_0(t_{e0})/dt^2) \approx B/(d^2s_0(t_e)/dt^2), \tag{33}$$

where $0\leq\lambda(B)\leq1$. The "approximately equal" sign shows that this equation may be used for an approximate correction of the observed timing of extremum $t_e$ to the extremum of the "proper" function $s_0(t)$. The "exactly equal" case will appear, if $d^3s_0(t)/dt^3=0$ at least in the interval between $t_e$ and $t_{e0}$.

These expressions can be similarly rewritten for a sum of the "proper" function $s_0(t)$ and an arbitrary trend $s_1(t)$, and determination of the root of the equation $d(s_0(t)+s_1(t))/dt=0$:

$$\delta = t_e - t_{e0} = \frac{\dot{s_1}}{\ddot{s_0} + \ddot{s_1}} \approx \frac{\dot{s_1}}{\ddot{s_0}} \tag{34}$$

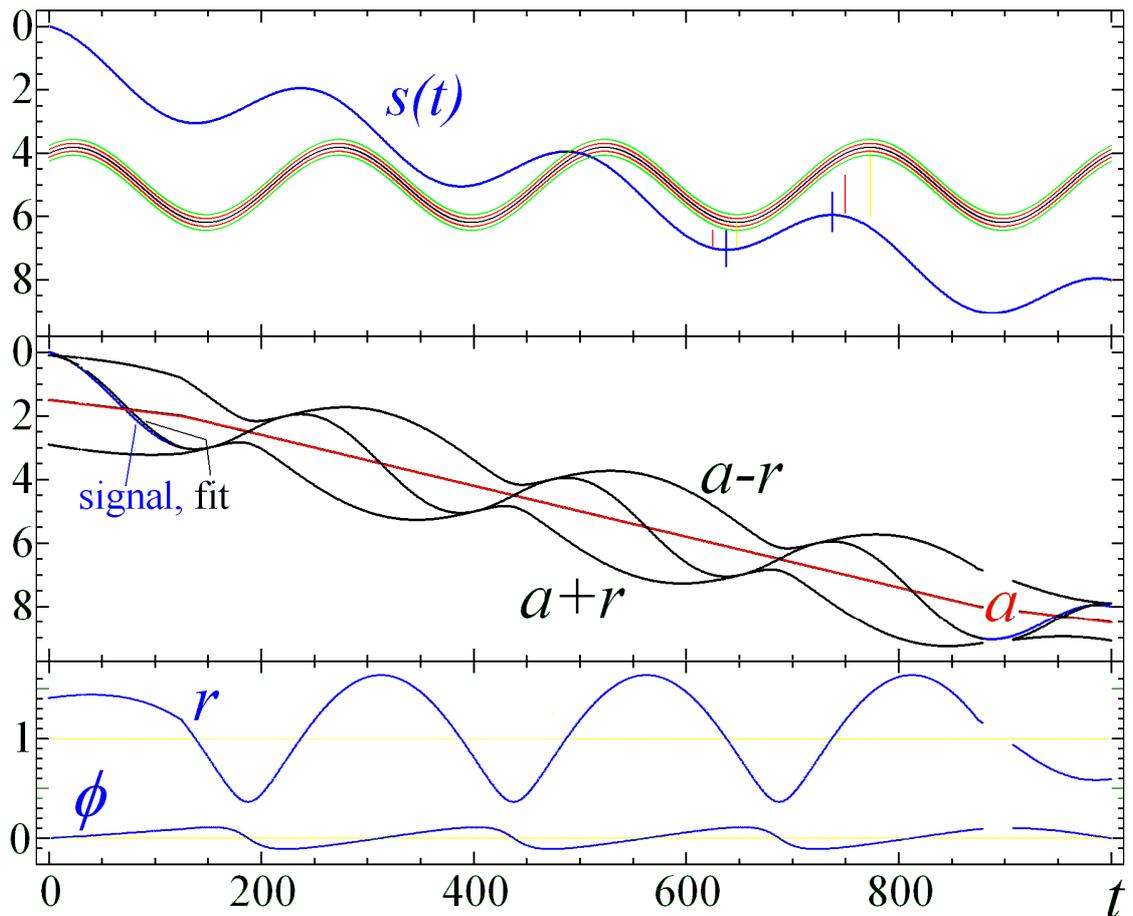

Fig. 4. Characteristics of the "global" sine (up) and "Running Sine" (middle and bottom) approximations for a test signal with a large trend. The positions of one minimum and one maximum are shown by vertical segments: for the model function $s(t)$ with trend (blue), without trend $s_0(t)$ (red), and for the "global" sine fit (yellow)

The accurate value can be determined again in some point $(t_{e0}+\lambda\delta)$ which is generally unknown and so is practically unusable. So one may use estimates $t_e$ and/or $t_{e0}$. The approximate expression is obviously becomes more accurate with decreasing ratio $|\ddot{s}_1/\ddot{s}_0|$.

This deviation $\delta$ is a property of functions with trends, and thus also of all approximations. Contrary to "global" approximation, where all parameters are not variable with time, in the "running sines" fit are present "waves". In our "crash test", the coincidence between the signal $s(t)$ and fit $\theta(t_0,\Delta t)$ is remarkable – exact in the middle, where the interval of smoothing is completely filled, and practically within line thickness close to the borders. Exact coincidence is between $a(t)$ and its theoretical expectation in the middle, and increases towards edges of the interval. However, the phase of the maximum $\phi$ shows an asymmetric wave of semi-amplitude 0.11. This is in an excellent agreement with a theoretical ("continuous" signal) value of

$$\phi = -(\arcsin(D/\pi))/2\pi, \tag{35}$$

which is equal to $-0.1098$ for our sample value $D=2$.

The argument of the "arcsin" function for the "running sine" approximation is exactly twice larger than for the function $s(t)$. This means that there is an absolute limit $|D|<\pi$ for applicability of the method. I.e. the maxima and minima will be present for $s(t)$, if $|D|<2\pi$, but for $\theta(t_0,\Delta t)$ only for $|D|<\pi$.

With an increasing $D$, the asymmetry increases, asymptotically tending to inclined segments and periodic jumps of phase by a step of unity.

### Application of the Method to Real Stars.

For an illustration, we apply the method to the same star AF Cyg, which was studied using the periodogram and scalegram analysis by Andronov and Chinarova (2012). 8738 observations from the AFOEV database have been used. The initial epoch $T_0=2453260.2$ and period $P_0=94.187^d$ were taken from a periodogram analysis and "global" fit (6). The observations and the parameters of the "running sine" approximation are shown in Fig.5. One may see variability of $a$, $r$, Max=$a$-$r$, min=$a$+$r$ and phase $\phi$. Most interesting is a significant trend at the beginning of the interval (t<52135. This may be interpreted as a longer period of ~181$^d$ instead of the basic one ~94$^d$. Four subsequent waves of long period are clearly seen at the dependence $a(t)$, what is expected for longer period $P>P_0$. There are some phase jumps by unity. They occur close to times when the semi-amplitude drops to zero. However, the phase restores (except the left interval), so one may interpret this as "lost" pulsation.

As this star is known to switch pulsation periods (see details in Andronov and Chinarova (2012)), one may justify effectivity of the method.

The mechanisms of pulsations of long-period pulsating stars were discussed e.g. by Kudashkina (2003), Fadeyev (2006), Samus' (2005). Observational phenomena detected using complementary methods of time series analysis should be used for more detailed classification and further theoretical models.

Historically, at first the method was applied to symbiotic stars UV Aur (Chinarova 1998) and V1329 Cyg (Chochol et al. 1999). Other examples of types of variability, where the method showed new results - intermediate polars MU Cam (Kim et al. 2005a) and BG CMi (Kim et al. 2005b), semi-regular pulsating variable RU And (Chinarova 2010), which switches between "Mira-type" large amplitude oscillations and time intervals of "constancy". Other similar stars of transition type were discussed by Marsakova and Andronov (2012). An interesting type of variability – "transient periodic oscillations" was detected in the magnetic dwarf nova DO Dra (Andronov et al. 2008), for which "running sines" may be recommended for an analysis.

**Final Remarks**

Contrary to the opinion listed in the epigraph, we propose to use a "net" of complementary methods for the time series analysis, which are statistically optimized for different types of variability. The method of "running sines" is one of such tools, which is related to the "running parabola" (Andronov 1997), "wavelet" (Andronov 1998, 1999), classic "running mean" and "least squares" (e.g. Whittaker and Robinson 1944) approximations.

As possible extensions of the method (see Andronov (2003) for a review on arbitrary test and weight functions), one may use e.g. a ("global or local") modification of $\Delta t$ (which is recommended to be $0.5P_0$), which may be determined to produce a best statistical accuracy of the smoothed signal (at specific points, or as an average over "discrete" or "continuous" interval), accuracy of (all or selected) timings or to maximize the "signal-to-noise" ratio. One may use smooth weight functions to avoid "jumps" in dependencies of parameters on time, like in the methods of "running parabola" or "wavelet". One may make a preliminary "pre-whitening" of data, removing the trend using any of the "low-frequency" approximations (algebraic and/or trigonometric polynomial etc.), and then finally just add the removed trend to $a(t)$.

A better approach is to use an approximation $s(t)= s_1(t) - R \cdot \cos(\omega(T-T_\mathrm{M}))$, where $s_1(t)$ describes periodic or a-periodic trend. However, a increasing number of parameters may lead to an increase of statistical errors, thus practically it may be recommended to use a linear trend (as in Eq. (28)) or a cosine term with much larger period. For small trends ($D \ll 1$), the model with linear trend will produce an increase of statistical errors by a factor of $(4/3)^{1/2} \approx 1.15$. Thus one could use a model with linear trend, if it is better than that without a trend. Similar option for the smoothing over time intervals (constant or inclined line) is included in the software MCV (Andronov and Baklanov 2004).

In some cases, the periods of stars are close to one year, and thus annual seasonal gaps in observations lead to bad coverage of the light curve, with subsequent relatively large error estimates. In these cases, we use a double value of $\Delta t = 1 \cdot P$, which obviously corresponds to twice worse time resolution when studying abrupt changes of these 3 characteristics. As the dependencies on $t_0$ with fixed $P$ and $\Delta t$ are generally discontinuous, it is recommended to make plots on a regular grid of $t_0$ instead of connecting them (and $\pm 1\sigma$ "error corridors") by lines.

Contrary to the "Morlet wavelet", the period is suggested to be constant (not dependent on $t_0$ and *not* proportional to $\Delta t$), and thus is to be determined preliminary using other methods.

For slow and small period variations, the dependence of the phase $\phi(t_0)$ (computed from $T_\mathrm{M}(t_0,\Delta t)$ for fixed period $P_0$ and chosen initial epoch $T_0$) is a scaled and translated version of the ("O-C") diagram (as phase $\phi = (O-C)/P$, and the epoch $E = \mathrm{int}((T-T_0)/P + 0.5)$.

Instead of sine $\varphi(t,t_0,\Delta t) = a - r \cdot \cos(\omega \cdot (t-T_M))$ with a free parameter, one may fix the time and $\varphi(t,t_0,\Delta t) = a - r \cdot \cos(\omega \cdot (t-t_0))$. In this case, there will be two parameters instead of 3. For sine signal $s(t) = A - R \cdot \cos(\omega \cdot (t-t_M))$, $a=A$, $r=R \cdot \cos(\omega \cdot (t_0-t_M))$ and the (local) maxima may be determined at maxima of $r$ (and minima for minima of $r$).

Similarly, one may use other waveforms $s(t) = A - R \cdot G( (t-t_M)/P_0)$, where $G(E)$ is some periodic function, and $\varphi(t,t_0,\Delta t) = a - r \cdot G( (t-t_0)/P_0)$. Then again the maxima of $r$ will occur at $t_0=t_M$, where $t_M$ is time of (local) maximum.

According to an effective number of observations used for the approximation, "Running sines" possess a place between "global" sine fit optimal for signals with good coherency (i.e. if the variations of phase $\phi$ during interval of observations are much smaller than unity), where all observations are taken into account, then a "wavelet fit" applicable for signals with possibly significant period changes at a scale of at least few periods, from one side, and "running parabola" (effective to changes from cycle to cycle) and then "running mean", from another side.

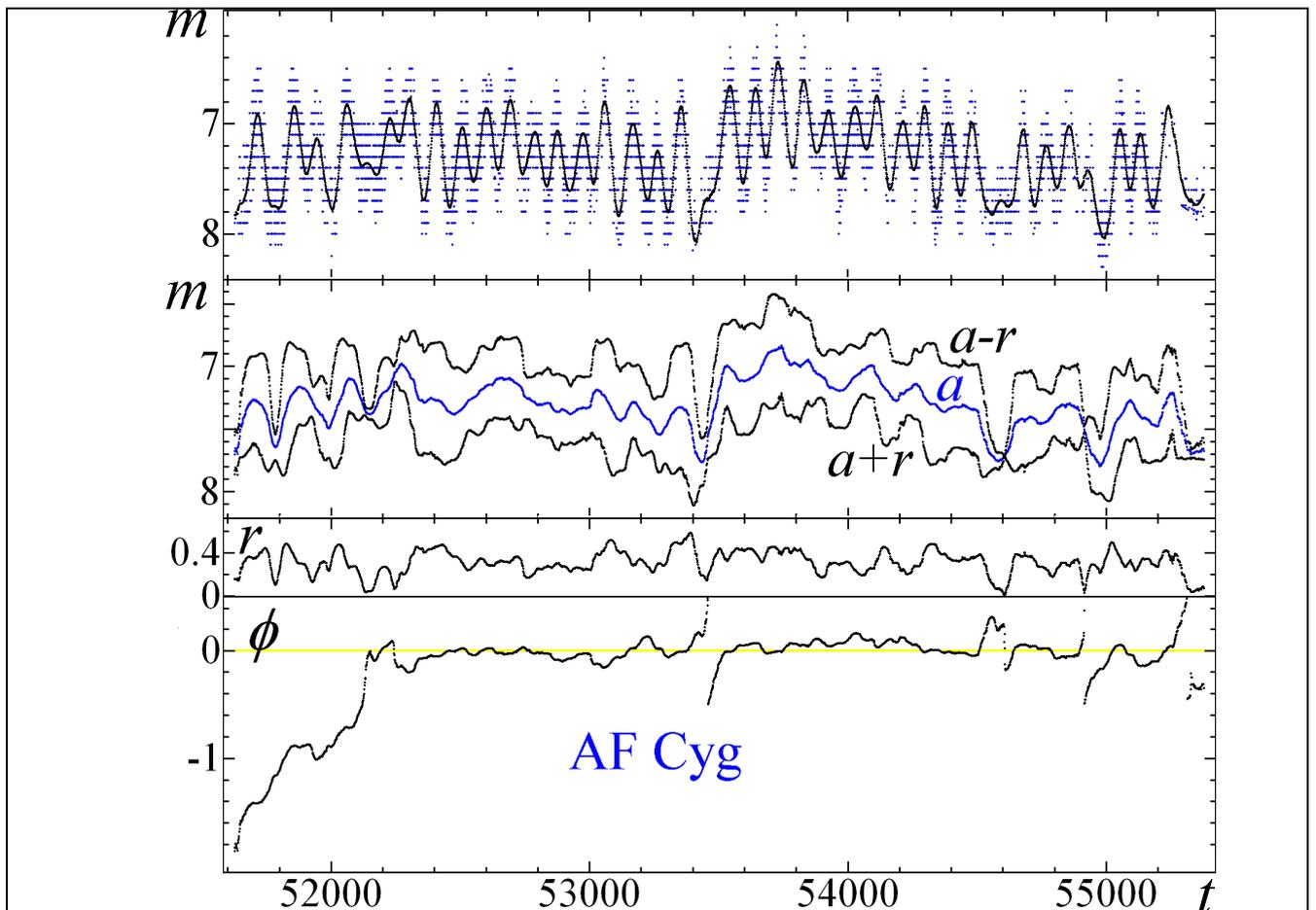

Fig. 5. Observations of semi-regular variable AF Cyg (up) and parameters of the "Running Sine" approximations (middle and bottom).

*Acknowledgements*. The authors are thankful to Dr. Bogdan Wszolek and the Institute of Physics of the Jan Dlugosz University for hospitality. The AFOEV database was used. This study is a part of the projects "Inter-Longitude Astronomy" (Andronov et al. 2010) and "Ukrainian Virtual Observatory" (Vavilova et al. 2012).